\documentclass[twocolumn]{aastex61}
\hypersetup{linkcolor=red,citecolor=cyan,filecolor=blue,urlcolor=black}

\newcommand{\hhco}{H$_2$CO}
\newcommand{\chhhoh}{CH$_3$OH}
\newcommand{\chhco}{H$_2$CCO}
\newcommand{\chhhcho}{CH$_3$CHO}
\newcommand{\chhhcn}{CH$_3$CN}
\newcommand{\chhhnc}{CH$_3$NC}
\newcommand{\chhhcch}{CH$_3$CCH}
\newcommand{\cch}{C$_2$H}
\newcommand{\ccchh}{C$_3$H$_2$}
\newcommand{\ccch}{C$_3$H}

\submitjournal{ApJ Letters}
\shorttitle{First detection of interstellar S$_2$H}
\shortauthors{Fuente et al.}

\begin{document}

\title{First detection of interstellar S$_2$H}

\correspondingauthor{Asunci\'on Fuente}
\email{a.fuente@oan.es}

\author[0000-0001-6317-6343]{Asunci\'on Fuente}
\affil{Observatorio Astron\'omico Nacional (OAN,IGN), Apdo 112, E-28803 
Alcal\'a de Henares, Spain}

\author{Javier R. Goicoechea}
\affiliation{Instituto de Ciencia de Materiales de Madrid (ICMM-CSIC), Sor Juana Inés de la Cruz, 3, E-28049
 Cantoblanco, Madrid, Spain}

\author{Jerome Pety}
\affiliation{Institut de Radioastronomie Millim\'etrique (IRAM),
300 rue de la Piscine, 38406 Saint Martin d'H\`eres, France}
\affiliation{LERMA, Observatoire de Paris, PSL Research University, CNRS, 
 Sorbonne Universit\'es, UPMC Univ. Paris 06, Ecole Normale Sup\'erieure, 
 F-75005 Paris, France.}

\author{Romane Le Gal}
\affiliation{Harvard-Smithsonian Center for Astrophysics, 60 Garden St., Cambridge, MA 02138, USA }

\author{Rafael Mart\'{\i}n-Dom\'enech}
\affiliation{Harvard-Smithsonian Center for Astrophysics, 60 Garden St., Cambridge, MA 02138, USA }

\author{Pierre Gratier}
\affiliation{Laboratoire d'Astrophysique de Bordeaux, Univ. Bordeaux, CNRS, 
B18N, all\'ee Geoffroy Saint-Hilaire, 33615 Pessac, France}

\author{Viviana Guzm\'an}
\affiliation{Joint ALMA Observatory (JAO), Alonso de C\'ordova 3107, Vitacura, Santiago, Chile}

\author{Evelyne Roueff}
\affiliation{LERMA, Observatoire de Paris, PSL Research University, CNRS, Sorbonne Universit\'es, UPMC Univ. Paris 06,  F-92190 Meudon, France}

\author{Jean Christophe Loison}
\affiliation{Institut des Sciences Mol\'eculaires de Bordeaux (ISM), CNRS, 
Univ. Bordeaux, 351 cours de la Lib\'eration, 33400, Talence, France}

\author{Guillermo M. Mu\~noz Caro}
\affiliation{Centro de Astrobiolog\'{\i}a (CSIC-INTA), Carretera de Ajalvir, 
km 4, Torrej\'on de Ardoz, 28850 Madrid, Spain}

\author{Valentine Wakelam}
\affiliation{Laboratoire d'Astrophysique de Bordeaux, Univ. Bordeaux, CNRS, 
B18N, all\'ee Geoffroy Saint-Hilaire, 33615 Pessac, France}

\author{Maryvonne Gerin}
\affiliation{LERMA, Observatoire de Paris, PSL Research University, CNRS, 
 Sorbonne Universit\'es, UPMC Univ. Paris 06, Ecole Normale Sup\'erieure, 
 F-75005 Paris, France.}

\author{Pablo Riviere-Marichalar}
\affiliation{Instituto de Ciencia de Materiales de Madrid (ICMM-CSIC), Sor Juana Inés de la Cruz, 3, E-28049
 Cantoblanco, Madrid, Spain}

\author{Thomas Vidal}
\affiliation{Laboratoire d'Astrophysique de Bordeaux, Univ. Bordeaux, CNRS, 
B18N, all\'ee Geoffroy Saint-Hilaire, 33615 Pessac, France}

\begin{abstract}  

We present the first detection of gas phase S$_2$H in the Horsehead, a moderately UV-irradiated nebula.
This confirms the presence of doubly sulfuretted species in the interstellar medium and opens a new challenge for sulfur chemistry.
The observed S$_2$H abundance is $\sim$5$\times$10$^{-11}$, only a factor 4-6 lower than
that of the widespread H$_2$S molecule. H$_2$S and S$_2$H are efficiently formed on the UV-irradiated icy grain mantles.
We performed ice irradiation experiments to determine the H$_2$S and S$_2$H photodesorption yields. The obtained values are 
$\sim$1.2$\times$10$^{-3}$  and  $<$1$\times$10$^{-5}$ molecules per incident photon for H$_2$S and S$_2$H, respectively. 
Our upper limit to the S$_2$H photodesorption yield suggests that photo-desorption is not a competitive mechanism to release
the S$_2$H molecules to the gas phase. Other desorption mechanisms such as chemical desorption, cosmic-ray desorption and grain 
shattering can increase the gaseous S$_2$H abundance to some extent. Alternatively, S$_2$H can be formed via gas phase reactions involving gaseous
H$_2$S and the abundant ions S$^+$ and SH$^+$. The detection of S$_2$H in this nebula could be therefore the result of the coexistence of 
an active grain surface chemistry and gaseous photo-chemistry.

\end{abstract}


\section{Introduction} 

Sulfur is one of the most abundant elements in the Universe (S/H$\sim$1.3$\times$10$^{-5}$)
and plays a crucial role in biological systems on Earth, so it is important to follow its chemical history in space.
Surprisingly, sulfuretted molecules are not as abundant as expected in the interstellar medium.
A few sulfur compounds have been detected in diffuse clouds demonstrating that the sulfur abundance
in these low density regions is close to the cosmic value \citep{Neufeld2015}. A moderate sulfur depletion 
(a factor of 4) is observed in the external layers of the photodissociation region (PDR) in the Horsehead nebula, as well 
\citep{Goicoechea06}. 
In cold molecular clouds, a large depletion of sulphur is usually considered to reproduce the observations 
(see for instance \citealp{Tieftrunk94}). 
However, recent models by \citet{Vidal17} explained the observational data without or with little sulfur 
depletion after updating the gas and grain chemistry.
In that case, HS and H$_2$S on the grains or atomic sulphur in the gas would contain most of the sulfur. In hot cores and corinos, 
\citet{Wakelam04} found that observations of S-bearing molecules would be better reproduced if sulfur was sublimated from grains 
in the atomic form or it is quickly converted into it. Thus far, the main solid or gas sulfur carrier is still debated.

With an adsorption energy of $\sim$1100 K \citep{Hasegawa93}, sulfur atoms are expected to stick on the
surfaces of grains with temperatures below $\sim$22~K. Here, because of the high hydrogen abundances and the
mobility of hydrogen in the ice matrix, sulfur atoms are expected to form H$_2$S. 
Indeed, gaseous H$_{2}$S is the most abundant S-bearing molecule in comets, with an abundance of up to 1.5\% relative to water 
\citep{bockelee00}.
A firm detection of H$_2$S in interstellar ices has not been reported yet.
A realistic upper limit of the H$_2$S abundance in interestellar ices is 1\% relative to water
that is 10 times lower than the cosmic abundance \citep{jimenez11}.
One possibility to explain the low fraction of H$_{2}$S in interstellar ices is that H$_2$S is processed 
by UV-photons or cosmic rays in the ice leading to the formation of 
other S-bearing species. OCS and tentatively SO$_2$ have been detected in icy 
mantles but their abundances are far too low 
to explain the missing S budget \citep{geballe85,palumbo95,boogert97}. 

Trying to provide new insights into the ice composition, 
experimental simulations of the irradiation of interstellar ices containing H$_{2}$S under astrophysically 
relevant conditions have been 
performed in laboratoty using UV photons \citep{jimenez11,jimenez14}, X-rays \citep{jimenez12}, 
or ions \citep{moore07,ferrante08,garozzo10}.
Energetic processing of H$_{2}$S-bearing ices readily generates sulfur-sulfur bonds, and 
the main S-bearing products in these experiments are H$_{2}$S$_{2}$ and S$_{2}$H that were
detected by \citet{jimenez11} through their infrared absorption bands. 
The molecule H$_{2}$S$_{2}$ could subsequently photodissociate forming S$_{2}$ and S$_{3}$ 
depending on the irradiation time. These molecules with two S atoms and even more 
could thus contain a significant fraction of the missing 
sulfur in dense clouds. In line with this work, \citet{Druard12} suggested that polysulphanes could be
a sulfur reservoir in the ice and are rapidly converted into atomic sulfur once in the gas phase.
\citet{Martin16} unsuccesfully searched for S$_2$H and H$_2$S$_2$ in the gas phase
toward the well-known hot corino, IRAS~16293$-$2422. 
The lack of gaseous S$_2$H and H$_2$S$_2$ was interpreted
as the consequence of the rapid destruction of these species once 
sublimated in such a warm and dense environment \citep{Martin16,Fort17}. 

In this Letter, we report the first interstellar detection of S$_2$H in the prototypical 
photo-dissociation region, the Horsehead. In Sect. 4, we discuss the possible grain-surface and gas-phase 
S$_2$H formation routes. New measurements of the photodesorption 
yields of S$_2$H and H$_2$S are presented in Sect. 5.

\section{Observations and data reduction} 
\label{sec:obs}
The data used in this work are from the Horsehead WHISPER  (Wide-band High-resolution Iram-30m Surveys at two 
Positions with Emir Receivers, PI: J. Pety) project and the Director's Discrectionary Time project D11-16. 
The Horsehead WHISPER project is a complete unbiased line survey of 
the 3, 2, and 1 mm bands using the IRAM 30m telescope. 
Two positions are observed: i) the HCO peak (RA=5$^{\rm h}$40$^{\rm m}$53$^{\rm s}$.936, 
Dec=−2$^\circ$28$'$00$''$, J2000), 
which is characteristic of the photo-dissociation region at the UV-illuminated surface of the Horsehead 
nebula \citep{Gerin09} (also referred to as PDR position), and ii) the DCO$^+$ peak (RA=5$^{\rm h}$40$^{\rm m}$55$^{\rm s}$.61, 
Dec=−2$^\circ$27$'$38$''$, J2000), 
which corresponds to a cold and UV-shielded condensation located less than 40$''$ away from the PDR edge
\citep{Pety07}. During the observations we used the
Position-Switching procedure with the reference position located at an offset ($-$100$''$,0)
relative to RA: 05$^{\rm h}$40$^{\rm m}$54$^{\rm s}$.27 Dec: $-$02$^\circ$28$'$00$''$.0.
Several lines of S$_2$H were tentatively detected towards the two positions observed in the WHISPER survey. 
In order to confirm the S$_2$H detection, we requested Director's Discrectionay 
Time (D11-16) to observe a single setup covering  the frequencies listed in Table 1. 
The merged S$_2$H spectra are shown in Fig. 1. Line intensities are given in main brightness
temperature (T$_{MB}$) and the lines were observed with a frequency resolution of 49 kHz.

In order to have a deeper insight into the S$_2$H chemistry, we compare the new S$_2$H observations with 
the H$_2$S 1$_{1,0}$$\rightarrow$1$_{0,1}$ map observed during April 2006 with the IRAM 30m telescope. 
These observations were done using the frequency switching mode and a spectral resolution of 40 kHz. 
Averaged noise level per resolution element at 168 GHz is rms($T_{MB}$)=170~mK.
The integrated intensity emission of the H$_2$S line varies 
between 1.0$-$1.5~K~km~s$^{-1}$ across the molecular cloud with an abrupt border in the west
(see Fig. 1). The H$_2$S 
emission presents a local minimum towards the DCO$^+$ peak, similar to the morphology observed in other
species such as CH$_3$OH \citep{Guzman11,Guzman13}, suggesting gaseous H$_2$S depletion towards this
cold dense core. 

\begin{table*}
\centering
\caption{Gaussian fits}
\begin{tabular}{lccccc}
\hline
\hline
Freq(MHz)   &   Area(K kms$^{-1}$)    &    v$_{lsr}$(km s$^{-1}$)  &   $\Delta$v(km s$^{-1}$)   &   T$_{MB}$(K)   &   rms (K) \\ \hline
\multicolumn{6}{l}{DCO$^+$ peak (core)} \\
94526.32     &    0.019 (0.002)  &   10.90 (0.05)  &   0.9 (0.1)  &  0.018  & 0.004 \\
94731.21     &    0.020 (0.003)  &   11.39 (0.13)  &   1.7 (0.2)  &  0.011  & 0.004 \\
110294.15    &    0.024 (0.004)  &   10.86 (0.05)  &   0.7 (0.1)  &  0.033  & 0.008  \\
110498.11    &    0.029 (0.004)  &   11.09 (0.06)  &   0.9 (0.1)  &  0.029  & 0.007  \\ \hline
\multicolumn{6}{l}{HCO peak (PDR)} \\
94526.32    &   0.016 (0.002)  &  10.97 (0.06)  &   0.9 (0.1)  &  0.018  &  0.004  \\
94731.21    &   0.023 (0.003)  &  11.03 (0.05)  &   0.9 (0.1)  &  0.023  &  0.004 \\
110294.15   &   0.026 (0.003)  &  10.87 (0.04)  &   0.7 (0.1)  &  0.037  &  0.008 \\
110498.11   &   0.019 (0.004)  &  11.07 (0.09)  &   0.9 (0.2)  &  0.020  &  0.007 \\
\hline
\end{tabular}
\end{table*}

\begin{figure*}
\centering
\includegraphics[width=13cm]{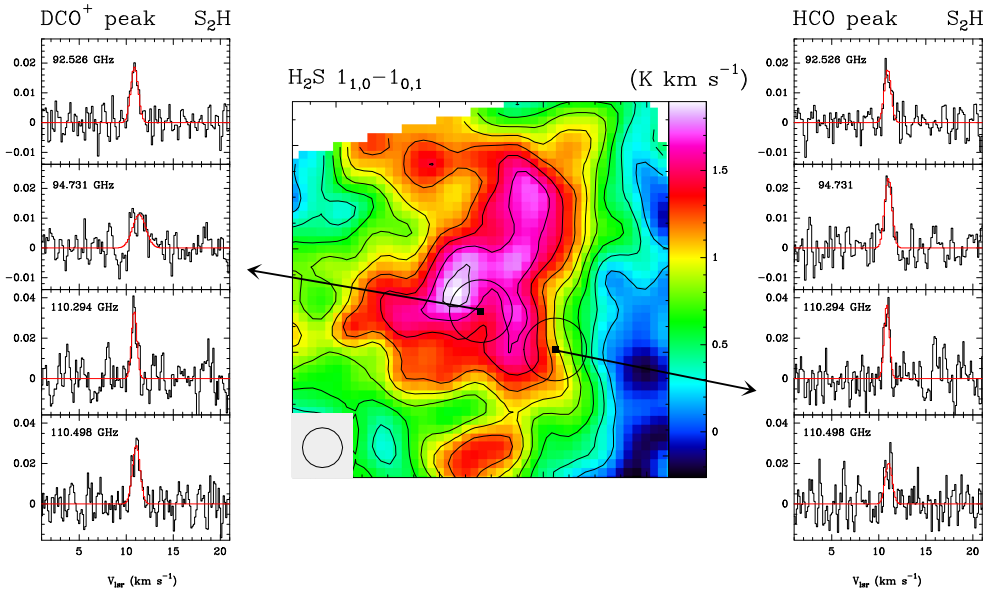}
\caption{In the central panel, we show the integrated intensity map of the H$_2$S 1$_{1,0}$$\rightarrow$1$_{0,1}$ line
(168.763 GHz). UV-illumination from $\sigma$Ori comes from the west (right). The beam is drawn in the bottom-left corner. 
Black circles around the surveyed positions indicate
the beam of the S$_2$H detections.
Spectra of the four S$_2$H lines detected towards the two positions targetted in the Whisper spectral are plotted in the left (DCO$^+$ peak) and right (HCO peak) panels. The frequency in GHz is indicated in the top-left corner. In red, the Gaussian fits shown in 
Table~1. }
\label{modelsfig}
\end{figure*}

\section{Column densities and abundances} 
\label{sec:s2h}

The rotational spectrum of S$_2$H was calculated by \citet{Tanimoto00}. The spectroscopic data 
can be found in the CDMS catalogue \citep{Muller00}. We have detected 
eight S$_2$H lines  located at 94526.1508, 94526.3208, 94731.0115, 94731.2080, 
110294.0282, 110294.1530, 110497.9666, and 110498.1104 MHz.
The S$_2$H hyperfine transitions are forming doublets very close in frequency ($\sim$0.12 MHz) that remain unresolved in
our data (see Fig.~1). In Table 1, we show the Gaussian fits to the observed line features, each one clearly detected with 
S/N$>$5. We have adopted as central frequency the one of the most intense component of the doublet. 
For this reason the central velocity shown in Table 1 is systematically shifted by $\sim$0.2$-$0.5 km s$^{-1}$ 
from the Horsehead systemic velocity, 10.5 km s$^{-1}$. 
We have checked possible contamination by other compounds using the CDMS and JPL catalogues. 
There is no other good candidate to be a carrier of these lines. The large linewidth of
the 94.731 GHz line towards the DCO$^+$ peak position, $\sim$1.7 km s$^{-1}$, is more likely due to the poor
baseline around this feature.

The WEEDS software has been used to simulate the S$_2$H spectrum
in the whole frequency coverage of the WHISPER survey  assuming LTE conditions.
We have fitted the detections and upper-limits of 97 lines with upper level energies lower than 75~K
found in the frequency range of the full survey using the Bayesian method described by \citet{Majumdar17}. 
The whole spectrum can be fitted assuming that the emission uniformly fills the beam  and the 
rotation temperatures and S$_2$H column densities listed in Table 1. The fitted line-widths are 
0.68$\pm$0.12 km s$^{-1}$ for the core and 0.63$\pm$0.1 km s$^{-1}$ for the PDR. 
The observed S$_2$H linewidths are consistent with the emission coming from the UV irradiated gas. 
Species that are more abundant in the cold and UV shielded gas of the core as DCO$^+$ and H$^{13}$CO$^+$, present 
narrower line-widths towards the DCO$^+$ peak than towards the PDR \citep{Goicoechea09}. 
However, others PDR-like species such as HCO present similar linewidths towards both positions \citep{Gerin09}.
This suggests that even towards the core position, the S$_2$H emission is mainly coming from the UV-illuminated 
layers of the cloud along the line of sight.

The S$_2$H rotation temperatures reveal 
subthermal excitation and are similar to those derived for other high dipole moment compounds
like o-H$_2$CO \citep{Guzman11}. The estimated abundance (wrt hydrogen nuclei) 
is $\sim$5$\times$10$^{-11}$ towards the DCO$^+$ peak and about a factor 
of 2 larger towards the HCO peak.

From the chemical point of view, it is interesting to compare the S$_2$H abundance with those of
the related species H$_2$S. Unfortunately there is only one transition of H$_2$S that
is easily observable with the 30m telescope given the physical conditions in the Horsehead, the 
o-H$_2$S 1$_{1,0}$$\rightarrow$1$_{0,1}$ line. Thus, we need to assume a rotation temperature to derive
the H$_2$S column density. Since the H$_2$S dipole moment ($\mu_b$=0.978 D; \citealp{Visvana1984}) 
is similar to those of S$_2$H  ($\mu_a$=1.161 D, $\mu_b$=0.827 D; \citealp{Peterson2008}), we assume 
the same rotation temperature for both molecules. With these assumptions and adopting an ortho-to-para ratio
of 3, we derive a H$_2$S abundance of $\sim$3$\times$10$^{-10}$ towards the two positions. 
This would imply that [S$_2$H]/[H$_2$S]=0.15$\pm$0.09 in the DCO$^+$ peak and 
[S$_2$H]/[H$_2$S]=0.27$\pm$0.14, toward the HCO peak. These numbers are consistent with the [S$_2$H]/[H$_2$S] ice
ratio obtained by \citet{jimenez12} in their simulations of UV irradiation of H$_2$S ices.
In the following, we qualitatively explore the possible surface and gas-phase formation routes
of S$_2$H.

\section{S$_2$H formation}

The formation of S$_2$H is a intricate problem due to the low H-SS energy bonding. 
Evidences for the formation of S$_2$H during irradiation of pure H$_2$S and H$_2$S:H$_2$O ice mixtures were provided 
by \citet{jimenez11} using the same experimental setup as the one described here.
One way of forming S$_2$H could be the grain surface 
reactions: s-H atom (hereafter, 's-' is used to refer to the solid phase)
addition on s-S$_2$, and s-S + s-HS reaction, followed by chemical desorption. However, 
the low exothermicity of the first reaction should prevent efficient chemical 
desorption \citep{Minissale16,Wakelam17}. 
The second reaction should not be efficient in cold cores because, below 15 K, S atom and HS 
radical are not mobile on ice considering the adsorption energies given by \citet{Wakelam17}. 
Moreover, S$_2$H is a very reactive 
species in the gas phase, reacting with H, N, C and O atoms without barrier, so likely also on surface. 
An alternative surface induced S$_2$H production may be 
s-H$_2$S$_2$ photo-dissociation-desorption: s-H$_2$S$_2$ + h$\nu$$\rightarrow$S$_2$H+H. s-H$_2$S$_2$ may be efficiently 
produced by the s-HS + s-HS reaction but it needs mobile HS on ice 
and so a high grain temperature. 

In the gas-phase, S$_2$H may be produced by the 
electronic dissociative recombination of H$_2$S$_2^+$. Even if there is no data on this reaction, the loss 
of one H atom is always an important exit channel on dissociative recombination \citep{Plessis10}. There are two 
known H$_2$S$_2^+$ production pathways: the S$^+$+H$_2$S$\rightarrow$H$_2$S$_2^+$+h$\nu$ reaction \citep{Anicich03}, 
despite the fact that the reference is an unpublished work and previous 
experimental studies did not identify this channel \citep{Smith81}, and 
the SH$^+$+H$_2$S$\rightarrow$H$_2$S$_2^+$+H reaction which is well characterized \citep{Anicich03}. 
We note that S$^+$ and SH$^+$ are only abundant in the UV-irradiated gas \citep{Gerin16}. Therefore, in spite of the large 
uncertainties in the reaction rates, we can conclude that these  
formation routes are only efficient in the UV-illuminated cloud surfaces.

\section{Experimental study of the photodesorption of S$_2$H and H$_2$S } 

\citet{jimenez11} showed that sulfur-sulfur bonds, in particular H$_2$S$_2$ and S$_2$H, are formed 
in irradiated ices. Here we focus on the determination of the S$_2$H and H$_2$S photodesorption yields which
are key to determine the origin (surface vs gas phase
chemistry) of the observed S$_2$H. For this aim, we performed experimental simulations under astrophysically relevant conditions 
using the ISAC setup \citep{mcaro10}, an ultra-high vacuum chamber with a work pressure on the order of 
4$\times$10$^{-11}$ mbar, corresponding to the pressure found in the interior of the pre-stellar cores. 
Sulfur is expected to be locked on the icy mantles in these regions, H$_2$S being the most abundant S-bearing molecule 
in cometary ices. Pure amorphous H$_2$S ice samples with thicknesses of about 40$\times$10$^{15}$ molecules cm$^{-2}$ 
were deposited from the gas phase (H$_2$S gas, Praxair, 99.8\%) onto a KBr substrate at 8 K, 
and subsequently irradiated using an F-type microwave-discharged hydrogen flow lamp with a vaccum-ultraviolet 
flux of 2$\times$10$^{14}$ photons cm$^{−2}$ s$^{-1}$ at the sample position \citep{mcaro10}. 
The emission spectrum of the lamp (reported in \citealp{Chen14}, and \citealp{Cruz14}) resembles 
that of the secondary UV field in dense cloud interiors, calculated by \citet{Gredel89}. 
A Pfeiffer Prisma quadrupole mass spectrometer (QMS) was used during irradiation of the ice samples to monitor the mass fragments 
$m/z$ = 34 (corresponding to photodesorbing H$_2$S molecules), and 
$m/z$ = 64 (corresponding to any desorbing photoproduct with a sulfur-sulfur bond, observed to form in \citet{jimenez12}). 
In our experiment, we did not monitore S$_2$H directly. However, if H$_2$S$_2$
or S$_2$H were desorbed, we would expect to detect all the fragments derived from these species, in particular
S$_2^+$. While photodesorption of H$_2$S was detected, no gaseous S$_2^+$ was observed (see Fig. 2).
The measured ion current was converted into a photodesorption yield following calibration of the QMS (see \citealp{Martin15}). 
Photodesorption of H$_2$S took place with a decreasing yield, reaching a steady-state value of 1.2$\times$ 10$^{-3}$ molecules per
incident photon after $\sim$30 minutes of irradiation, which corresponds to the fluence experienced by 
ice mantles during the typical cloud lifetime \citep{Shen04}. A factor of 2 is assumed as the error in the photodesorption yield values due to 
the uncertainties in the calibration process, see \citealp{Martin16b}. Following the non-detection of any sulfur-sulfur photoproduct, 
an upper limit of 1$\times$10$^{-5}$ molecules per incident photon (the sensitivity limit of our QMS) was assumed for the photodesorption of S$_2$H. 
Direct S$_2$H photodesorption or H$_2$S$_2$ photo-dissociation-desorption are therefore not expected to be the origin of the 
gaseous S$_2$H.

\begin{centering}
 \begin{figure}
  \includegraphics[width=9cm]{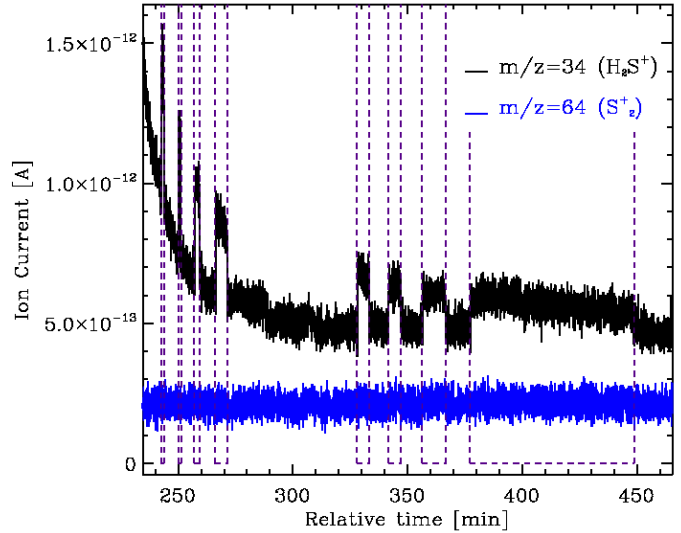}
  \caption{Photodesorption of H$_2$S (black) detected by the QMS during irradiation of a pure H$_2$S ice sample. No increase of the measured 
  ion current for the mass fragment $m/z$ = 64 (blue, corresponding to any sulfur-sulfur photoproduct) was detected. 
  Irradiation intervals are indicated with vertical dashed lines. Signals are shifted for clarity.}
 \end{figure}
\label{exp}
\end{centering}

\begin{table*}
\caption{Summay of column densities and fractional abundances}
\begin{center}
\begin{tabular}{lc ccc|ccc}
\hline
\hline
Molecule    & \multicolumn{1}{c}{} & \multicolumn{3}{c|}{DCO$^+$ peak} &  \multicolumn{3}{c}{HCO peak} \\
 & HPBW     & T$_{rot}$  &  N(X)  &  N(X)/N$_H$
            & T$_{rot}$  &  N(X)  &  N(X)/N$_H$ \\
 & ($''$)   &  (K) &  (cm$^{-2}$)  &   
            &  (K) &  (cm$^{-2}$)  &   \\ \hline \hline
H$_2$       &  12  &      &  2.9$\times$10$^{22}$   & 0.5   &   & 1.9$\times$10$^{22}$   & 0.5  \\ 

S$_2$H      &  22 - 26    & 8.73$_{-1.10}^{+1.36}$   & 3.0$_{-0.6}^{+0.9}$$\times$10$^{12}$  
&  5.2$_{-1.0}^{+1.5}$$\times$10$^{-11}$  & 12.69$_{-1.54}^{+1.78}$  & 3.3$_{-0.7}^{+1.2}$$\times$10$^{12}$ &
8.7$_{-1.9}^{+3.1}$$\times$10$^{-11}$  \\

H$_2$S$^1$      &  14         & 9$_{-1}^{+1}$  & 1.9$_{-0.3}^{+0.4}$$\times$10$^{13}$  &  3.3$_{-0.6}^{+0.7}$$\times$10$^{-10}$  &
12$_{-2}^{+2}$  & 1.2$_{-0.2}^{+0.1}$$\times$10$^{13}$  &    3.1$_{-0.5}^{+0.3}$$\times$10$^{-10}$ \\

\hline
\end{tabular}
\end{center}

\noindent
{\scriptsize
(1) We assume the rotation temperatures derived from S$_2$H. \\
}

\end{table*}


\begin{table}
\caption{Column density upper limits}
\begin{center}
\begin{tabular}{lrrr}
\hline
\hline
Molecule    &  \multicolumn{1}{c}{Freq} & \multicolumn{1}{c}{rms$^1$} & \multicolumn{1}{c}{N$_X^2$ }  \\
   &  \multicolumn{1}{c}{(GHz)} & \multicolumn{1}{c}{(mK)} & \multicolumn{1}{c}{(cm$^{-2}$)}  \\ \hline \hline
S$_2$H$_2$       &  139.885  & 9   & $<$8.5$\times$10$^{11}$  \\                       
HSO              &  158.391  & 30  & $<$1.5$\times$10$^{12}$  \\                      
HO$_2$           &  130.260  & 11  & $<$4.7$\times$10$^{11}$  \\   
H$_2$O$_2$       &   90.365  &  4  & $<$1.0$\times$10$^{12}$  \\ 
\hline
\end{tabular}
\end{center}

\noindent
{\scriptsize
(1) The rms has been calculated for a channel width of $\approx$0.3 km s$^{-1}$. 
The obtained rms is similar in the two
surveyed positions. (2) 3$\sigma$ upper limits assuming LTE, 
T$_{rot}$=10 K and a linewidth of 0.6 km s$^{-1}$}

\end{table}

\section{Discussion and conclusions} 

At a distance of 400 pc, the Horsehead is a PDR viewed nearly edge-on and illuminated  by  the  O9.5V  
star $\sigma$Ori at a  projected  distance of $\sim$3.5 pc. The intensity of the incident FUV radiation 
field is $\chi$=60 relative to the interstellar radiation field in Draine units.
This PDR presents a differentiated chemistry from others associated with nearby HII regions such as 
the Orion Bar. One main difference is that the dust temperature is around $\sim$20-30 K in the PDR
\citep{Goicoechea09}, i.e. below or close to the 
sublimation temperature of many species, allowing a rich surface chemistry on the irradiated surfaces. 
Our unbiased line survey has provided valuable hints on the chemistry of this region. 
The detection of the molecular ions CF$^+$ and HOC$^+$ towards the HCO peak are well
understood in terms of gas-phase photochemistry \citep{Guzman12}. We learned that there is 
  an efficient top-down chemistry in the PDR, in which large
  polyatomic molecules or small grains are photo-destroyed into smaller
  hydrocarbon molecules/precursors, such as {\cch}, {\ccchh}, {\ccch}~and
  C$_3$H$^+$~\citep{Pety12,Guzman15}. The detection of several complex 
organic molecules (COMs) towards the warm ($T_\mathrm{kin}\sim60$\,K) PDR and its associated cold
($T_\mathrm{kin}\sim20$\,K) core was unexpected. In fact, the chemical complexity reached in the Horsehead is extraordinarily
  high with COMs of up to 7 atoms: HCOOH, \chhco, \chhhcho~and {\chhhcch}~\citep{Guzman14}.
  Current pure gas-phase models cannot reproduce the inferred {\hhco}, {\chhhoh} and COMs
  abundances in the Horsehead PDR~\citep{Guzman11,Guzman13}, which supports the grain surface origin of these molecules.
  \citet{Gal17} was able to reproduce the observed COMs abundances using a chemical model with grain surface chemistry
  and found that chemical desorption, instead of photodesorption, is the main process to release COMs to the gas phase.
  {\chhhcn} and {\chhhnc}, key species for the formation of prebiotic
  molecules, seem to have a very specific formation pathway in the PDR~\citep{Gratier13}. 
 The Horsehead is therefore an excellent site 
 to study the influence of UV radiation on the grain surface chemistry and its subsequent impact
 on the gas phase. 

We present the first detection of S$_2$H in the Horsehead.  The observed S$_2$H abundance is $\sim$5$\times$10$^{-11}$, 
only a factor 4-6 lower than
that of H$_2$S. Our laboratory experiments show that the H$_2$S and S$_2$H photodesortion yields 
are 1.2$\times$10$^{-3}$ and  $<$1$\times$10$^{-5}$ molecules per incident photon, respectively. 
Although S$_2$H can be formed on warm (T$_d$$>$15 K) grains, 
our upper limit to the S$_2$H photodesorption yield suggest that this mechamism is not efficient to release the S$_2$H molecules
from the grain mantles. Other desorption mechanisms such as chemical desorption, cosmic-ray desorption and grain 
shattering could increase the S$_2$H abundance in gas phase. 
S$_2$H can also be formed in gas-phase by reactions involving H$_2$S and the ions S$^+$ and SH$^+$. These ions are expected to
be abundant in the external layers of the PDR \citep{Goicoechea06}. The photodesorption
of H$_2$S could hence boost the S$_2$H production in gas phase. We conclude that the abundance of S$_2$H in the Horsehead is more likely
the consequence of the favorable physical conditions prevailing in this nebula where grain mantles irradiated by UV photons coexist with 
the ions S$^+$ and SH$^+$ that are only abundant in PDRs. 

One interesting issue is to compare the sulfur and oxygen chemistry. We have not detected H$_2$S$_2$, HSO, H$_2$O$_2$ and
HO$_2$ in the Horsehead with the upper limits shown in Table~3.  
We find interesting that the column densities of HSO and HO$_2$ are lower than that of S$_2$H, although the oxygen elemental abundance 
is 30 times greater than that of sulfur.  
In gas phase, S$_2$H is mainly formed through S$^+$ + H$_2$S$\rightarrow$H$_2$S$_2^+$ + h$\nu$ and 
SH$^+$ + H$_2$S$\rightarrow$H$_2$S$_2^+$ + H followed by dissociative recombination of H$_2$S$_2^+$. 
Oxygen and sulfur have indeed similar reactivity but, due to their different ionization potentials, 
O$^+$ is expected less abundant than S$^+$ and then the O$^+$ and OH$^+$ reactions play a smaller role. 
We have also compared the SH$^+$ + H$_2$O and SH$^+$ + H$_2$S gas-phase reactions which may be intermediate 
paths at work for producing SOH and S$_2$H, respectively. The channel towards HSO$^+$  + H$_2$ reaction is endothermic
in opposite to the channels towards S$_2$H$^+$ + H$_2$ and S$_2$H$_2^+$ + H. Therefore, in gas phase the formation of S$_2$H
is favored relative to HSO. HSO and related species have not been observed in space thus far \citep{Caz16,Fort17}.
Laboratory experiments demonstrate that grain surface chemistry involving H$_2$O and H$_2$S also present different pathways.  
Photo-desorption experiments reported by \citet{Cruz17} show that
H$_2$O$_2$ is not formed in UV irradiated water ice. In contrast, \citet{jimenez11} showed that H$_2$S$_2$ is formed when a
H$_2$S and H$_2$S-H$_2$O ices are irradiated, providing a path to form species with two sulfur atoms. 
Summarizing, sulfur and oxygen are not analogues in the gas-phase and surface chemistry, and the 
comparison of their related species requires the full chemical modelling of the region.

\acknowledgements
We thank the Spanish MINECO for funding support from
AYA2016-75066-C2-1/2-P, AYA2012-32032 and ERC under ERC-2013-SyG, G. A. 610256 NANOCOSMOS.
This work was supported by the Programme National $``$Physique et Chimie du Milieu Interstellaire$''$
(PCMI) of CNRS/INSU with INC/INP co-funded by CEA and CNES.
     
\bibliographystyle{aasjournal}

\begin{thebibliography}
\expandafter\ifx\csname natexlab\endcsname\relax\def\natexlab#1{#1}\fi
\providecommand{\url}[1]{\href{#1}{#1}}

\bibitem[Anicich, V.G. (2003)]{Anicich03} 
Anicich, V. G., 2003, JPL Publication 2003, 03-19 NASA 

\bibitem[Bockel\'ee-Morvan et al.(2000)]{bockelee00} 
Bockel\'ee-Morvan, D., Lis, D.~C., Wink, J.~E., et al. 2000, A\&A, 353, 1101

\bibitem[Boogert et al.(1997)]{boogert97} 
Boogert, A.~C.~A., Schutte, W.~A., Helmich, F.~P., Tielens, A.~G.~G.~M.,\& Wooden, D.H. 1997, A\&A, 317, 929

\bibitem[Cazzoli et al.(2016)]{Caz16} Cazzoli, G., Lattanzi, V., Kirsch, T., et al.\ 2016, A\&A, 591, A126 

\bibitem[Chen et al.(2014)]{Chen14} Chen, Y.-J., Chuang, K.-J., Mu{\~n}oz Caro, G.~M., et al.\ 2014, ApJ, 781, 15 

\bibitem[Cruz-Diaz et al.(2014)]{Cruz14} Cruz-Diaz, G.~A., Mu{\~n}oz Caro, G.~M., Chen, Y.-J., \& Yih, T.-S.\ 2014, A\&A, 562, A119 

\bibitem[Cruz-Diaz et al.(2017)]{Cruz17} Cruz-Diaz, G.~A., Mart{\'{\i}}n-Dom{\'e}nech, R., Moreno, E., Mu{\~n}oz Caro, G.~M., \& Chen, Y.-J.\ 2017, arXiv:1711.05679 

\bibitem[Druard \& Wakelam(2012)]{Druard12} Druard, C., \& Wakelam, V.\ 2012, MNRAS, 426, 354 

\bibitem[Ferrante et al.(2008)]{ferrante08} 
Ferrante, R.~F., Moore, M.~H., Spiliotis, M.~M., \& Hudson, R.~L. 2008, ApJ, 684, 1210 

\bibitem[Fortenberry \& Francisco(2017)]{Fort17} Fortenberry, R.~C., \& Francisco, J.~S.\ 2017, ApJ, 835, 243 

\bibitem[Garozzo et al.(2010)]{garozzo10} 
Garozzo, M., Fulvio, D., Kanuchova, Z., Palumbo, M.~E., \& Strazzulla, G. 2010, A\&A, 509, A67 

\bibitem[Geballe et al.(1985)]{geballe85} 
Geballe, T.~R., Baas, F., Greenberg, J.~M., \& Schutte, W. 1985, A\&A, 146, L6 

\bibitem[Gerin et al.(2009)]{Gerin09} Gerin, M., Goicoechea, J.~R., Pety, J., \& Hily-Blant, P.\ 2009, A\&A, 494, 977 

\bibitem[Gerin et al.(2016)]{Gerin16} Gerin, M., Neufeld, D.~A., \& Goicoechea, J.~R.\ 2016, \araa, 54, 181 

\bibitem[Goicoechea et al.(2006)]{Goicoechea06} Goicoechea, J.~R., Pety, J., Gerin, M., et al.\ 2006, A\&A, 456, 565 

\bibitem[Goicoechea et al.(2009)]{Goicoechea09} Goicoechea, J.~R., Compi{\`e}gne, M., \& Habart, E.\ 2009, ApJl, 699, L165 

\bibitem[Gredel et al.(1989)]{Gredel89} Gredel, R., Lepp, S., Dalgarno, A., \& Herbst, E.\ 1989, ApJ, 347, 289 

\bibitem[Gratier et al.(2013)]{Gratier13} Gratier, P., Pety, J., Guzm{\'a}n, V., et al.\ 2013, A\&A, 557, A101 

\bibitem[Guzm{\'a}n et al.(2011)]{Guzman11} Guzm{\'a}n, V., Pety, J., Goicoechea, J.~R., Gerin, M., \& Roueff, E.\ 2011, A\&A, 534, A49 

\bibitem[Guzm{\'a}n et al.(2012)]{Guzman12} Guzm{\'a}n, V., Pety, J., Gratier, P., et al.\ 2012, A\&A, 543, L1 

\bibitem[Guzm{\'a}n et al.(2013)]{Guzman13} Guzm{\'a}n, V.~V., Goicoechea, J.~R., Pety, J., et al.\ 2013, A\&A, 560, A73 

\bibitem[Guzm{\'a}n et al.(2014)]{Guzman14} Guzm{\'a}n, V.~V., Pety, J., Gratier, P., et al.\ 2014, Faraday Discussions, 168, 103 

\bibitem[Guzm{\'a}n et al.(2015)]{Guzman15} Guzm{\'a}n, V.~V., Pety, J., Goicoechea, J.~R., et al.\ 2015, ApJl, 800, L33 

\bibitem[Hasegawa \& Herbst(1993)]{Hasegawa93} Hasegawa, T.~I., \& Herbst, E.\ 1993, MNRAS, 261, 83 

\bibitem[Jim\'enez-Escobar \& Mu\~noz Caro(2011)]{jimenez11} 
Jim\'enez-Escobar, A. \& Mu\~noz Caro, G.~M. 2011, A\&A, 536, A91 

\bibitem[Jim\'enez-Escobar et al.(2012)]{jimenez12} 
Jim\'enez-Escobar, A., Mu\~noz Caro, G.~M., Cicarelli, A., Cecchi-Pestellini, C., Candia, R., \& Micela, G. 2012, ApJL, 751, L40 

\bibitem[Jim\'enez-Escobar et al.(2014)]{jimenez14} 
Jim\'enez-Escobar, A., Mu\~noz Caro, G.~M., \& Chen, Y.-J. 2014, MNRAS, 443, 343 

\bibitem[Le Gal et al.(2017)]{Gal17} Le Gal, R., Herbst, E., Dufour, G., et al.\ 2017, A\&A, 605, A88 

\bibitem[Majumdar et al.(2017)]{Majumdar17} Majumdar, L., Gratier, P., Andron, I., Wakelam, V., \& Caux, E.\ 2017, MNRAS, 467, 3525 

\bibitem[Mart{\'{\i}}n-Dom{\'e}nech et al.(2015)]{Martin15} Mart{\'{\i}}n-Dom{\'e}nech, R., Manzano-Santamar{\'{\i}}a, J., Mu{\~n}oz Caro, G.~M., et al.\ 2015, A\&A, 584, A14 

\bibitem[Mart{\'{\i}}n-Dom{\'e}nech et al. (2016a)]{Martin16} Mart{\'{\i}}n-Dom{\'e}nech, R., 
Jim{\'e}nez-Serra, I., Mu{\~n}oz Caro, G.~M., et al.\ 2016, A\&A, 585, A112 

\bibitem[Mart{\'{\i}}n-Dom{\'e}nech et al.(2016b)]{Martin16b} Mart{\'{\i}}n-Dom{\'e}nech, R., Mu{\~n}oz Caro, G.~M., \& Cruz-D{\'{\i}}az, G.~A.\ 2016, A\&A, 589, A107 

\bibitem[Minissale et al.(2016)]{Minissale16} Minissale, M., Dulieu, F., Cazaux, S., \& Hocuk, S.\ 2016, A\&A, 585, A24 

\bibitem[Moore et al.(2007)]{moore07} 
Moore, M.~H., Hudson, R.~L., \& Carlson, R.~W. 2007, Icarus, 189, 409 

\bibitem[M{\"u}ller et al.(2005)]{Muller00} 
M{\"u}ller, H.~S.~P., Schl{\"o}der, F., Stutzki, J., \& Winnewisser, G.\ 
2005, J. Mol. Struct., 742, 215 

\bibitem[Mu{\~n}oz Caro et al.(2010)]{mcaro10} Mu{\~n}oz Caro, G.~M., Jim{\'e}nez-Escobar, A., Mart{\'{\i}}n-Gago, 
J.~{\'A}., et al.\ 2010, A\&A, 522, A108 

\bibitem[Neufeld et al.(2015)]{Neufeld2015} Neufeld, D.~A., Godard, B., 
Gerin, M., et al.\ 2015, A\&A, 577, A49 

\bibitem[Palumbo et al.(1995)]{palumbo95} 
Palumbo, M.~E., Tielens, A.~G.~G.~M., Tokunaga, A.~T. 1995, ApJ, 449, 674 

\bibitem[Plessis et al.(2010)]{Plessis10} 
Plessis, S., Carrasco, N. \& Pernot, P., 2010, J. Chem. Phys., 133, 13 

\bibitem[Peterson et al.(2008)]{Peterson2008} 
 Peterson, K.A., Mitrushchenkov, A., \& Francisco, J.S., 2008, Chem. Phys. 346, 34.

\bibitem[Pety et al.(2007)]{Pety07} Pety, J., Goicoechea, J.~R., Hily-Blant, P., Gerin, M., \& Teyssier, D.\ 2007, A\&A, 464, L41 

\bibitem[Pety et al.(2012)]{Pety12} Pety, J., Gratier, P., Guzm{\'a}n, V., et al.\ 2012, A\&A, 548, A68 

\bibitem[Shen et al.(2004)]{Shen04} Shen, C.~J., Greenberg, J.~M., Schutte, W.~A., \& van Dishoeck, E.~F.\ 2004, A\&A, 415, 203 

\bibitem[Smith et al.(2004)]{Smith81} Smith, D., Adams, N. G. \& Lindinger, W., 1981, The Journal of Chemical Physics, vol 75, n 7, 3365 

\bibitem[Tanimoto et al.(2000)]{Tanimoto00} 
Tanimoto, M., Klaus, T., M{\"u}ller, H.~S.~P., \& Winnewisser, G. 
2000, J. Mol. Spectrosc., 199, 73 

\bibitem[Tieftrunk et al.(1994)]{Tieftrunk94} Tieftrunk, A., Pineau des Forets, G., Schilke, P., \& Walmsley, C.~M.\ 1994, A\&A, 289, 579 

\bibitem[Vidal et al.(2017)]{Vidal17} Vidal, T.~H.~G., Loison, J.-C., Jaziri, A.~Y., et al.\ 2017, MNRAS, 469, 435 

\bibitem[Viswanathan et al.(1984)]{Visvana1984}
R. Viswanathan \& T. R. Dyke, 1984, J. Mol. Spectrosc. 103, 231.

\bibitem[Wakelam et al.(2004)]{Wakelam04} Wakelam, V., Caselli, P., Ceccarelli, C., Herbst, E., \& Castets, A.\ 2004, A\&A, 422, 159 

\bibitem[Wakelam et al.(2017)]{Wakelam17} Wakelam, V., Loison, J.-C., Mereau, R., \& Ruaud, M.\ 2017, Molecular Astrophysics, 6, 22 


\end{thebibliography}

\end{document}